\documentclass[twocolumn,showpacs,preprintnumbers,amsmath,amssymb]{revtex4}

\usepackage[dvips]{graphicx}
\usepackage{dcolumn}
\usepackage{bm}

\usepackage{color}

\newcommand{\beq}{\begin{equation}}
\newcommand{\eeq}{\end{equation}}

\begin{document}
\def\bfB{\mbox{\bf B}}
\def\bfQ{\mbox{\bf Q}}
\def\bfD{\mbox{\bf D}}
\def\etal{\mbox{\it et al}}

\title{Nonlinear dynamos at infinite magnetic Prandtl number}
\author{Alexandros Alexakis}

\address{Laboratoire de Physique Statistique de l'Ecole Normale
Sup\'erieure, UMR CNRS 8550, 24 Rue Lhomond, 75006 Paris Cedex 05, France.}

\date{\today}

\begin{abstract}

The dynamo instability is investigated in the limit of infinite magnetic Prandtl number.
In this limit the fluid is assumed to be very viscous so that
the inertial terms can be neglected and the flow is slaved to the forcing.
The forcing consist of
an external forcing function that drives the dynamo flow and the 
resulting Lorentz force caused by the back reaction of the magnetic field.  
The flows under investigation are the Archontis flow, and the ABC flow forced at two different scales.
The investigation covers roughly three orders of magnitude of the magnetic Reynolds number above onset.
All flows show a weak increase of the averaged magnetic energy as the magnetic Reynolds number is increased. 
Most of the magnetic energy is concentrated in flat elongated structures that 
produce a Lorentz force with small solenoidal projection so that the resulting magnetic 
field configuration was almost force-free.
Although the examined system has zero kinetic Reynolds number  at sufficiently large magnetic Reynolds number 
the structures are unstable to small scale fluctuations that result in a chaotic
temporal behavior.
%

\end{abstract}


\maketitle

Dynamo instability is considered to be the main mechanism for the generation 
and sustainment of magnetic field throughout the universe \cite{Zeldovich}.
In this scenario an initially small magnetic field is amplified by currents induced solely
by the motion of a conducting fluid.  This process saturates when the magnetic field becomes
strong enough for the Lorentz force to act back on the flow and reduce its ability 
to further amplify the magnetic energy.
The exact effect the Lorentz force has on the flow and which property of the flow 
is altered to prevent further magnetic field amplification is 
the subject of many current investigations.

One can in principle envision many scenarios for saturation especially in the presence 
of turbulence.
For example, saturation can occur because the magnetic  field suppresses the fluid flow thus reducing the magnetic Reynolds number of 
the flow to its critical value.  This behavior is expected close to the dynamo onset for laminar flows.
In another scenario, 
dynamo can saturate by suppressing the chaotic stretching of the magnetic field lines
\cite{Cattaneo1,Mak};
or the dynamo can saturate 
because the folding of the field lines is modified to be less
constructive.
%
Finally, saturation of the dynamo can be due to correlations between
the velocity and the magnetic field without suppressing the ability of the 
flow to amplify an uncorrelated passive vector field 
thus in principle still be a dynamo flow (see for example \cite{Cattaneo2,Axel,Martin}). 

%

In principle there is no unique answer to this question
and the list above does not exhaust all possibilities.
The exact mechanism can be a combination of different processes
whose choice can depend on the type of forcing and the examined parameter range.
Identifying the mechanisms in simple flows can help
in unveiling the  dependence of the amplitude of magnetic field saturation
on the magnetic Reynolds number $R_m$ and the fluid Reynolds number $Re$.
This dependence is important for astrophysics, 
bearing in mind that both numbers are very large,
with $1\ll Rm \ll   Re$ for the Sun and $1\ll Re \ll Rm$ for the galaxy.

In this large parameter space, it is interesting to examine special limits 
of the governing equations, in order to obtain a basic understanding of the
non-linear dynamo behavior. This is attempted  
in this work. Here saturated dynamos are investigated in the 
simplified limit of large viscosity. 
In this system the inertial terms can be neglected 
and the resulting flow is a Stokes flow driven by the external force
and the Lorentz force. 
This limit is typically used as a model for the small scale galactic dynamo 
for which the magnetic Prandtl number 
(which is equal to the ratio of the viscosity to the magnetic diffusivity) is very large.
In previous investigations \cite{Scheko1,Scheko2,Scheko3,Scheko4}  the forcing was assumed random and changing  
in time in order to model turbulent fluctuations at the viscous cut-off scale.
The turbulent fluctuations at this scale result in the fastest growing dynamo modes  
at the kinematic stage of the dynamo.  
Here the opposite limit is examined and a time independent forcing is used.
A steady forcing might not be suitable in describing the turbulent velocity fluctuations,
however it allows to investigate the reaction of the magnetic field to
large scale flows with long time correlations that could be more relevant in the saturated stage.
Another interesting property of this limit is that the system is stripped off
all hydrodynamic instabilities. Thus, when comparing with results of turbulent MHD flows
one can distinguish which effect is due to turbulence and hydrodynamic instabilities
and which solely due to the effect of the Lorentz force.

This paper is structured as follows, the next section
describes the dynamical equations in the limit considered.
Section II explains the numerical method and its limitations.
Section III describes the results for each of the examined flows.
Section IV gives some common scaling laws observed for all flows,
and conclusions are drawn in the last section.

\section{Equations and set up}

We consider flows in a triple periodic domain of size $L=2\pi$ driven by
a simple time independent body force. The non-dimensional MHD equations for a unit density fluid are
then given by
\begin{eqnarray}
{Re}(\partial_t {\bf u} + {\bf u \cdot \nabla u}) & =& {\bf b \cdot \nabla b} -\nabla P + \nabla^2 {\bf u} + {\bf F}  \label{NS}\\
\partial_t {\bf b} + {\bf u \cdot \nabla b}& =& {\bf b \cdot \nabla u}           +\frac{1}{R_m} \nabla^2 {\bf b}.    \label{MF}
\end{eqnarray} 
${\bf F}$ is the external body force that in the absence of magnetic fields sustains a 
velocity field ${\bf u}$, of unity root mean squared value $U=\langle {\bf u^2} \rangle^{1/2}=1.0$
(where $\langle \cdot \rangle $ stands for spatial average).
The Reynolds number is defined as $Re= U L/(2\pi \nu)$ where $\nu$ is the kinematic viscosity
and we are going to consider the limit $Re \ll 1$.
The magnetic field is expressed by 
${\bf b} = \sqrt{Re} \, \tilde{\bf b}/U$, where $\tilde{\bf b}$ is the dimensional
magnetic field measured in units of velocity.
The factor of $\sqrt{Re}$ has been introduced
to set the Lorentz force (that leads to saturation)
to the same order of magnitude as the forcing term.
The magnetic Reynolds number is defined as
$R_m=U L/(2\pi \eta)$, where $\eta$ is the magnetic diffusivity.
$P$ is the pressure that ensures the incompressibility condition $\nabla \cdot {\bf u}=0$. 

We are interested in the limit $Re\ll 1$ while $R_m$ is still finite;
thus the magnetic Prandtl number $P_m$, defined as the ratio of the two Reynolds numbers $P_m \equiv Re/R_m=\nu/\eta$, 
tends to infinity.
In this limit then, the left hand side of equation \ref{NS} can be dropped
and velocity field is slaved to the forcing and the Lorentz force and is given by
\beq
{\bf u} = - \nabla^{-2} \left[ {\bf b \cdot \nabla b} -\nabla P + {\bf F} \right] ,
\label{ST}
\eeq
where $\nabla^{-2}$ stands for the inverse laplacian. Eq. \ref{ST} represents the Stokes-flow of a viscous fluid
under the influence of the body-force ${\bf F}$ and the solenoidal projection of the Lorentz force $ {\bf b \cdot \nabla b} -\nabla P$. 
The approximations made here, are valid in the small Reynolds number limit provided that the velocity 
gradients do not become too sharp or the velocity time scale too short.  
In the examined calculations the velocity field appears to have a smooth behavior both in time and space
as $R_m$ is increased, so this condition is not violated. 

Equations \ref{MF} and \ref{ST} form a closed set of equations that are under investigation in this work.
They are going to be referred to as the infinite magnetic Prandtl number equations. 
The evolution equation for the magnetic energy is then given by:
\beq
\frac{1}{2} \partial_t \langle b^2 \rangle = \langle {\bf b \cdot (\nabla u) \cdot b} \rangle - \frac{1}{R_m} \langle (\nabla b)^2 \rangle 
\label{ener1}
\eeq
Multiplying eq. \ref{ST} by $\nabla^2{\bf u}$ space averaging and subtracting from eq. \ref{ener1} we obtain
\beq
\frac{1}{2} \partial_t \langle b^2 \rangle =
      \langle {\bf u \cdot F } \rangle  - \langle (\nabla u)^2 \rangle  - \frac{1}{R_m} \langle (\nabla b)^2 \rangle .
\label{ener2}
\eeq
Eq. \ref{ener2} is the energy balance equation that expresses that the rate of change of magnetic energy
is equal to the rate energy is injected to the system by $ \langle {\bf u \cdot F } \rangle$ minus the energy lost
by the viscous damping $\epsilon_u \equiv \langle (\nabla u)^2 \rangle$ and Ohmic dissipation  $\epsilon_b \equiv \langle (\nabla b)^2 \rangle/R_m$.
It worth pointing out that even in the limit $R_m\to \infty$ energy is dissipated at all scales 
due to the viscous term.
%
It is also noted here that the kinetic energy of the flow does not enter the energy balance equation since ${\bf u}$
is a slaved vector field. If we did add it, it would be of order $Re$ smaller than the magnetic energy.
In the remaining text we will refer to the kinetic and magnetic energy as $E_u \equiv \frac{1}{2}\langle {\bf u^2} \rangle$
and $E_b \equiv \frac{1}{2}\langle {\bf b^2} \rangle$ respectably, although for a true ``dimensional" comparison between the two a factor of 
$Re$ should be included ({\it ie } $E_{tot}=E_u+E_b/Re$).

The infinite magnetic Prandtl equations  conserve one ideal invariant (for $R_m=\infty$): the magnetic helicity \cite{Moffatt69}.
Multiplying equation \ref{MF} by the vector potential ${\bf a}$ (such that ${\bf b = \nabla \times a}$) and space averaging we obtain 
\beq
\frac{1}{2} \partial_t \langle {\bf a \cdot b } \rangle = -\frac{1}{R_m} \langle {\bf b\cdot \nabla \times b} \rangle,
\eeq 
the magnetic helicity $H_m \equiv \frac{1}{2}\langle {\bf a \cdot b } \rangle$ is thus conserved for infinite $R_m$. 
For large but finite $R_m$ this invariance is expected to be violated at scales $\ell \sim 1/\sqrt{R_m}$. 
Since there is no large scale source term for magnetic helicity, it can be generated or destroyed only at these small scales.
In dynamo investigations the initial magnetic field is very small and thus also the initial
magnetic helicity is negligible. However a flow with positive (negative) hydrodynamic helicity 
will generate  negative (positive) magnetic helicity in the larger scales where it will ``pile-up"
and positive (negative) helicity in the small scales where it will be dissipated.  
Thus although the flow itself does not generate helicity it drives preferentially one sign of helicity
to the small scales where it is dissipated.
So if the saturated state is magnetically helical with one sign of helicity 
it implies that the opposite sign of helicity has been dissipated at the small scales.

\section{Numerical method}
The infinite magnetic Prandtl number equations \ref{MF},\ref{ST}  were solved in a triple periodic
domain of size $L=2\pi$ using a standard pseudo-spectral method and a third order in time 
Runge-Kuta.
The code was based on a full MHD code \cite{Pablo1,Pablo2}. The truncated fields were 
dialliased every time a quadratic nonlinear term was calculated based on the 2/3 rule.

The resolution used varied from $64^3$ up to $256^3$ grid points. With a resolution of $256^3$ one 
can typically achieve magnetic Reynolds numbers of the order of a few hundreds for the kinematic problem
depending on the flow.
However, in the non-linear regime, 
    the velocity amplitude  is significantly reduced (in some cases by two orders of magnitude)
and the magnetic energy is dissipated also by the viscous term 
making the flow less efficient at generating small scales. 
Thus, although at the examined resolution only magnetic Reynolds numbers of few hundreds can be achieved
for the kinematic problem, for the nonlinear problem magnetic Reynolds numbers of a few thousand can easily be achieved 
if the initial conditions were also at the nonlinear stage.
The following procedure was used for the numerical runs  that were performed:
Starting from a well resolved kinematic dynamo simulation (very small initial magnetic field) the system
was evolved until non-linear saturation was achieved. The output of this run was then used 
as initial conditions for the next run at a higher magnetic Reynolds number. In many cases a  random 
perturbation was also added in the initial conditions.
A run was considered well-resolved if the maximum of the current density spectrum was sufficiently larger than
the value of the current density spectrum at the smallest resolved scale (typically by an order of magnitude).  
 
The strongest limitation in these runs turned out to be the time of integration. Although large magnetic Reynolds numbers
can be be achieved in this limit, with growth rates of the order of the inverse turnover time, the magnetic field 
with large Prandtl number has much longer memory time than a typical turbulent flow. Thus to achieve saturation for saturation
each run has to be integrated for time scales of the order of the diffusion time.   
  
\section{The examined flows}

Three different forcing functions $F$ were used to generate the dynamo flows.
The first two are members of the ABC family: 
\begin{eqnarray}
u_x = A \sin(k_f z) + B \cos(k_f y), \nonumber  \\
u_y = C \sin(k_f x) + A \cos(k_f z),            \\
u_z = B \sin(k_f y) + C \cos(k_f x). \nonumber
\end{eqnarray}
The ABC flows have been extensively investigated in the literature \cite{ABC_h1,ABC_h2,ABC_h3},
especially for their relevance in dynamo theory \cite{ABC_m1,ABC_m2,ABC_m3,ABC_m4}. 
They are fully-helical satisfying ($\nabla \times {\bf u =}- k_f{\bf  u}$).
Here the case for $A=B=C=1/\sqrt{3}$ is investigated for $k_f=1$ and
for the case $k_f=2$.    

The third flow under investigation is a variation of the ABC family
where only the cosine terms are kept:
\begin{eqnarray}
u_x =  A\cos(k_f y), \nonumber  \\
u_y =  B\cos(k_f z),            \\
u_z =  C\cos(k_f x). \nonumber 
\end{eqnarray}
This flow was first investigated in \cite{Arc_0,Arc_1} and was shown to result in
dynamo with large saturation amplitudes of the magnetic field \cite{Arc_2,Arc_3,Arc_4,Arc_5,Ponty}. 
This flow will be referred as the Archontis flow.  
Unlike the ABC flows, it has zero global helicity.
In the performed runs in this work have $A=B=C=\sqrt{2/3}$ and $k_f=1$.

\section{Results}

\subsection{Archontis Flow}

\begin{figure}
\includegraphics[width=8cm]{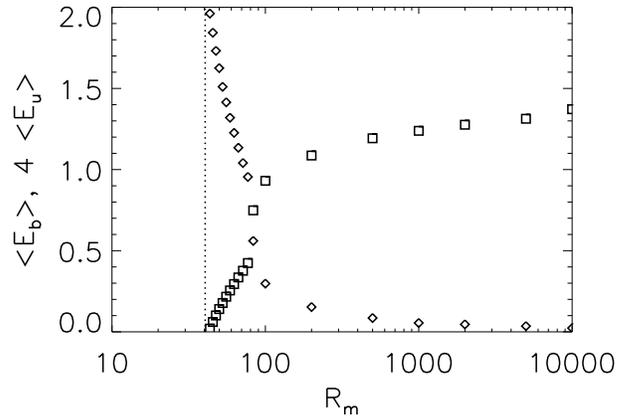}
\caption{\label{fig1} Time averaged magnetic (squares) and kinetic (diamonds) energy
                      as a function of $R_m$ for the Archontis flow.  }
\end{figure}

We begin with the non-helical Archontis flow.
Figure \ref{fig1} shows the saturation values
of the magnetic energy $E_b=\frac{1}{2}\|{\bf b}\|^2$ (squares) and
of the kinetic  energy $E_u=\frac{1}{2}\|{\bf u}\|^2$ (diamonds)
as a function of the magnetic Reynolds number $R_m$.
For small values of $R_m$ there is no dynamo and thus $\|{\bf b}\|^2=0$ and $\|{\bf u}\|^2=1$.
The onset of the dynamo instability is at $R_{mc}\simeq 42.5$. For small deviations from this
critical value the magnetic energy grows as $E_b\sim (Rm-R_{Mc})$ as a normal mode expansion
would predict \cite{Fauve,Petrelis}, and the kinetic energy sharply decreases. The dynamo is stationary at this stage.
As $R_m$ is further increased
the system goes through a series of bifurcations that we do not fully explore here.
There is an other critical value $R_{mc}'\simeq 80$ after which
the evolution of the magnetic energy appears to be chaotic.
We avoid using the term turbulence since the flow is in the small $Re$ regime, however
many common features with turbulence like power-law scalings are present in this regime.
It is noted that this chaotic state does not result from a random forcing or
hydrodynamic turbulence but it is solely generated by magnetic instabilities.
The dynamo remains in this chaotic state
for larger $R_m$ for all the examined runs. Magnetic field energy continues to grow
as $R_m$ is further increased. This increase however appears to be logarithmic $E_b\sim \ln(R_m)$
or possibly as a weak power law $E_n \sim R_m^\delta$ with $\delta<0.15$.
On the other hand, the kinetic energy is decreased approaching zero as $R_m$ is increased.

\begin{figure}
\includegraphics[width=8cm]{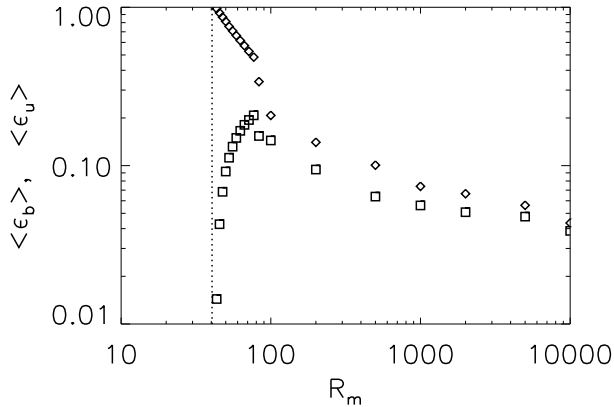}
\caption{\label{fig2} Time averaged Ohmic (squares) and viscous (diamonds) dissipation rates
                      as a function of $R_m$ for the Archontis flow. }
\end{figure}

Figure \ref{fig2} shows the Ohmic (squares) and the viscous (diamonds) dissipation rate as a function
of $R_m$. As the dynamo threshold is crossed  the viscous
dissipation rate is decreased and the Ohmic dissipation rate is increased.
The total dissipation rate is smaller than the laminar no-dynamo dissipation rate
({\it ie } the system is less dissipative when the dynamo is turned on).
The increase of the Ohmic dissipation rate continues until the critical value $R_{mc}'$,
where the chaotic behavior starts. After this value both the viscous and
the Ohmic dissipation rate are decreasing and become almost equal. This decrease continued up until the highest
examined magnetic Reynolds number. Here, the results of the numerical simulations, leave
open the possibility that at infinite $R_m$ a zero-injection, zero-dissipation state
could be reached.

\begin{figure}
\includegraphics[width=8cm]{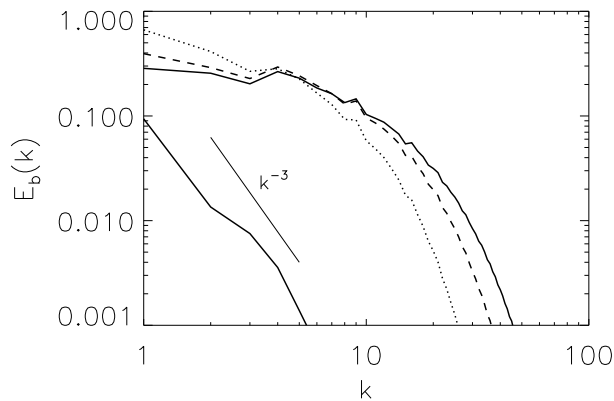}
\caption{\label{fig3} Magnetic energy spectrum for three values of $R_m$:
          $R_m=2000$ (dotted), $R_m=5000$ (dashed), $R_m=10000$ (solid)
          for the Archontis flow.
          The lower solid line shows the kinetic energy spectrum for $R_m=10000$. }
\end{figure}

Finally, in figure \ref{fig3} the energy spectra
for the three largest examined values of $R_m$, are shown.
As expected the cut-off wave number due to magnetic diffusivity is increased
as $R_m$ is increased.
What is also observed is that the bulk of the magnetic energy also
moves from the large to the small scales as the magnetic Reynolds number is increased,
with the slope of the spectrum changing from a negative value to almost flat.
It should also be noted that the large scale magnetic energy is decreasing as $R_m$ is increased
while the total magnetic energy is increased.

\begin{figure}
\includegraphics[width=6cm]{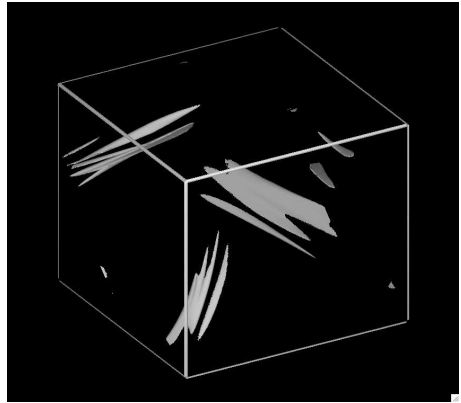}
\caption{\label{fig4} Magnetic energy density isosurfaces, for the Archontis Flow, $R_m=10^4$.
The surfaces correspond to 50\% of the maximum magnetic energy density or 10 times the average
magnetic energy.}
\end{figure}

Figure \ref{fig4} shows magnetic energy density isosurfaces, for the Archontis Flow, $R_m=10^4$.
Most of the energy is consecrated in elongated almost two-dimensional structures.  
These structures have been reported before in kinematic dynamo simulations
\cite{Arc_2} and are referred to as ``magnetic ribbons". Although these structures occupy a small fraction
of the total volume they are responsible for almost bringing to a halt the flow
in the entire domain.

\subsection{ABC Flow, $k_f=1$}

Next we examine the helical ABC flow forced at wavenumber $k_f=1$.
This is perhaps the most examined flow in periodic boxes for dynamo action.
The onset of instability and the kinematic regime of the dynamo has been
examined in the works \cite{ABC_m1,ABC_m2,ABC_m3}. It has been shown that the first onset
of the dynamo is at $R_{mc1} \simeq 15.5$. This mode however stops being a dynamo 
when $R_m$ becomes larger than $R_{mc2} \simeq 31$.
A second unstable mode appears at $R_{mc3}=43$ and the dynamo instability is present
for all larger values of $R_m$.

\begin{figure}
\includegraphics[width=8cm]{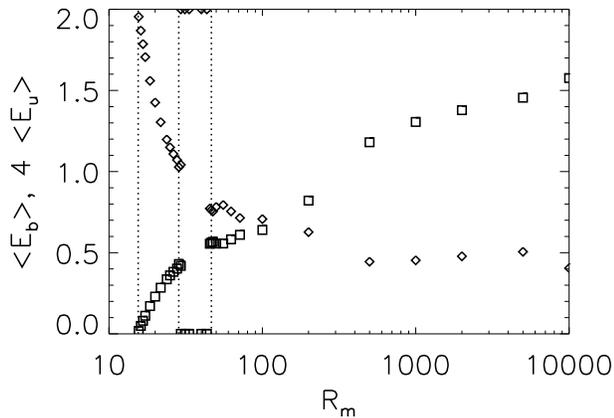}
\caption{\label{fig5} Time averaged magnetic (squares) and kinetic (diamonds) energy
                      as a function of $R_m$ for the ABC, $k_f=1$ flow. }
\end{figure}

Figure \ref{fig5} shows the saturation levels of the magnetic (squares) and kinetic (diamonds) energy
as a function of the magnetic Reynolds number. The dynamo instability begins
at $R_{mc1}$ as has been previously found and it exhibits a supercritical
bifurcation. At the no-dynamo window between $R_{mc2}$ and $R_{mc3}$ the
instability is sub-critical at both ends of the window. This is revealed by
the finite amplitude of saturation of the magnetic energy right at the onset 
of the instability. The extend of the sub-criticality appears however to be very 
small: Non-linear solutions with finite magnetic energy were found 
inside the no-dynamo window but very close to the boundaries of the window.   

For larger values of $R_m$ the magnetic energy continues to grow
and this flow  also exhibits chaotic behavior.
As in the Archontis Flow 
the magnetic energy 
shows a logarithmic (or a weak power law) increase with $R_m$. Unlike
the Archontis flow however the kinetic energy saturates to a finite 
value $E_u\simeq 0.125$. This is mostly due to the oscillatory nature
of the ABC dynamo that is driving a time dependent flow in the non-linear regime.

\begin{figure}
\includegraphics[width=8cm]{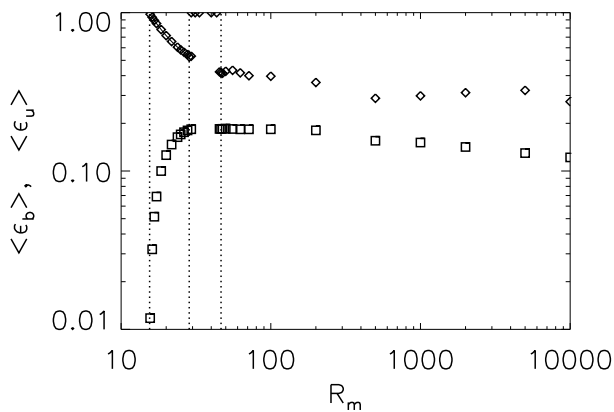}
\caption{\label{fig6} Time averaged Ohmic (squares) and viscous (diamonds) dissipation rates
                      as a function of $R_m$ for the ABC, $k_f=1$ flow.}
\end{figure}

The Ohmic (squares) and viscous (diamonds) energy dissipation for the ABC dynamo is
shown in figure \ref{fig6} as a function of $R_m$.
After the initial increase of the Ohmic dissipation at larger values of
$R_m$, the two dissipations become almost independent on $R_m$
with the viscous dissipation being twice larger than the Ohmic dissipation.
As in the previously examined flow, in the presence of dynamo the system is less dissipative.

\begin{figure}
\includegraphics[width=8cm]{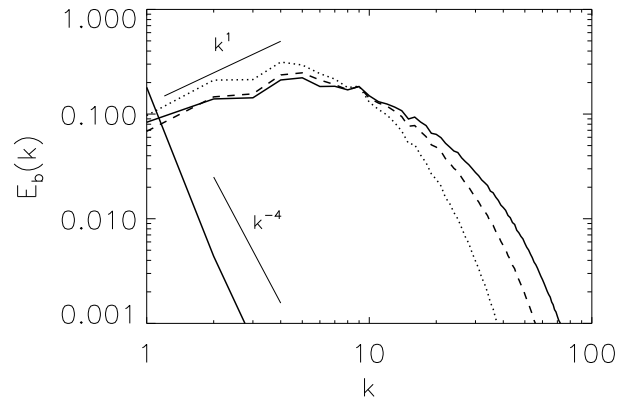}
\caption{\label{fig7} Magnetic energy spectrum for three values of $R_m$:
          $R_m=2000$ (dotted), $R_m=5000$ (dashed), $R_m=10000$ (solid)
          for the ABC, $k_f=1$ flow.
          The lower solid line shows the kinetic energy spectrum for $R_m=10000$. }
\end{figure}

The magnetic energy spectra 
shown in figure \ref{fig7} appear to have a positive slope 
at small wave numbers. 
This is in contrast with the MHD simulations with order one Prandtl number for which
at the saturated state most of the magnetic energy is concentrated close to
the forcing scale \cite{pablo3,pablo4}.
The positive slope extends to larger wavenumbers as the 
magnetic Reynolds number is increased. 

\begin{figure}
\includegraphics[width=6cm]{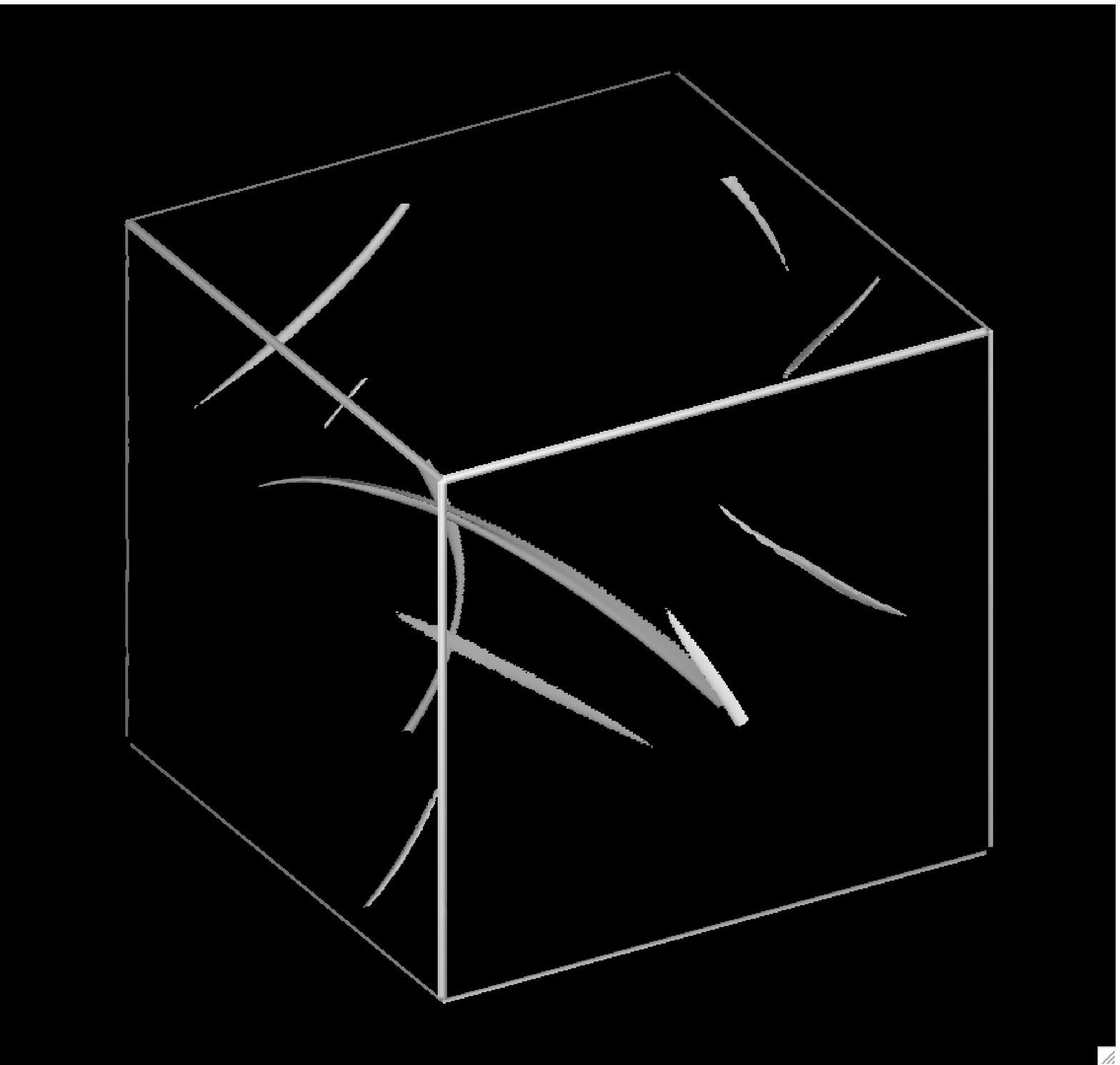}
\caption{\label{fig8} Magnetic energy density isosurfaces, for the ABC Flow $k_f=1$, $R_m=10^4$.
The surfaces correspond to 50\% of the maximum magnetic energy density or 10 times the average
magnetic energy.}
\end{figure}

\begin{figure}
\includegraphics[width=6cm]{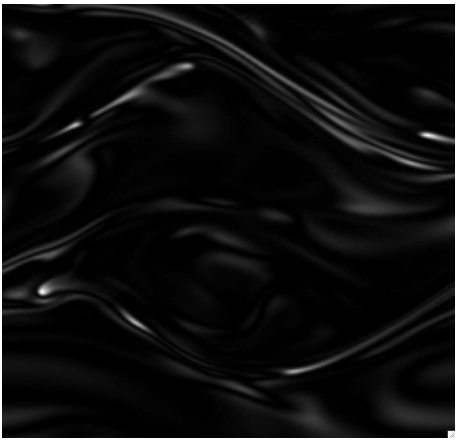}
\caption{\label{fig9} Grey-scale image of magnetic energy through
the mid plane $z=\pi$ of figure \ref{fig8}, (ABC Flow $k_f=1$, $R_m=10^4)$.
Bright regions indicate large magnetic energy.} 
\end{figure}

The magnetic energy is consecrated in elongated structures as shown in figure \ref{fig8}.
A grey-scale image of magnetic energy through the mid plane of the box shown in fig.\ref{fig9} 
indicates more clearly that these structures are also flat and part of sheet-like structures
that intensify in regions of strong shear; much like the the ``ribbons" observed in the Archontis flow.


\subsection{ABC flow, $k_f=2$}

Finally we examined the case for which the ABC flow is
forced at wave number $k_f=2$. The difference with the $k_f=1$ case
is that more magnetic modes are present in the system 
that move the onset of the instability to smaller values, and 
allow magnetic helicity to concentrate in the large scales.
 
\begin{figure}
\includegraphics[width=8cm]{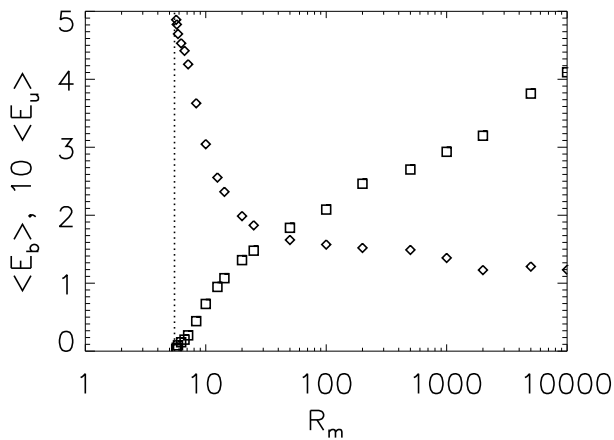}
\caption{\label{fig10} Time averaged magnetic (squares) and kinetic (diamonds) energy
                      as a function of $R_m$ for the ABC, $k_f=2$ flow.}
\end{figure}

\begin{figure}
\includegraphics[width=8cm]{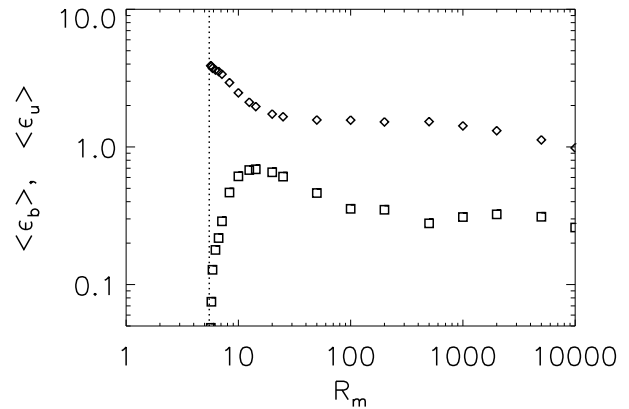}
\caption{\label{fig11} Time averaged Ohmic (squares) and viscous (diamonds) dissipation rates
                       as a function of $R_m$ for the ABC, $k_f=2$ flow. }
\end{figure}

Figure \ref{fig10} shows the magnetic (squares) and kinetic (diamonds) energy
at saturation for this flow. 
The onset of the instability appears at $R_m\simeq 6$ and there is no
no-dynamo window as in the $k_f=1$ case.
For large values of $R_m$ 
the magnetic energy can be seen to increase at least as fast as the logarithm of the magnetic Reynolds number.
This increase is more pronounced than the previously examined flows.
The kinetic energy saturates at a finite value roughly equal to one fifth of the value in the absence of dynamo.  
The two dissipation rates Ohmic (squares) and viscous (diamonds)
shown in figure \ref{fig11}, appear to be proportional
for sufficiently large $R_m$ with the Ohmic dissipation
rate being roughly four times smaller than the 
viscous dissipation. 

\begin{figure}
\includegraphics[width=8cm]{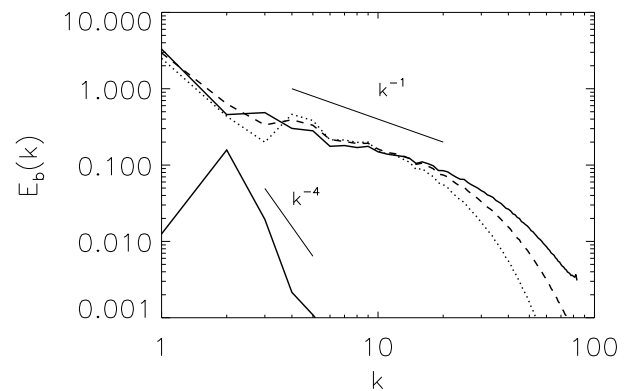}
\caption{\label{fig12} Magnetic energy spectrum for three values of $R_m$:
          $R_m=2000$ (dotted), $R_m=5000$ (dashed), $R_m=10000$ (solid)
          for the ABC, $k_f=2$ flow.
          The lower solid line shows the kinetic energy spectrum for $R_m=10000$.  }
\end{figure}

The most pronounced difference from the previous cases
can be seen in the magnetic energy spectra (Fig. \ref{fig12}).
Unlike the previously examined cases the bulk of the magnetic energy
is concentrated at the large scales as a result of the inverse
cascade of the magnetic helicity \cite{hely_1,hely_2,hely_3,hely_4,hely_5,hely_6}.
In the small scales the energy spectrum appears to follow a 
power law behavior $E_b(k)\sim k^\alpha$ with an exponent close to
$\alpha = -1$.


\begin{figure}
\includegraphics[width=8cm]{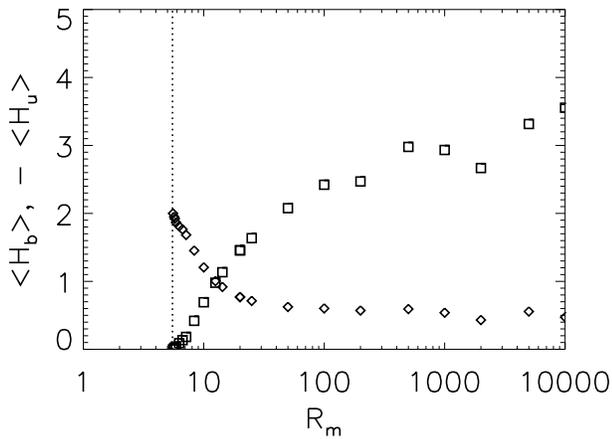}
\caption{\label{fig13} Time averaged magnetic (squares) and kinetic (diamonds) helicity as a function of $R_m$ 
for the ABC, $k_f=2$ flow.}
\end{figure}

\begin{figure}
\includegraphics[width=8cm]{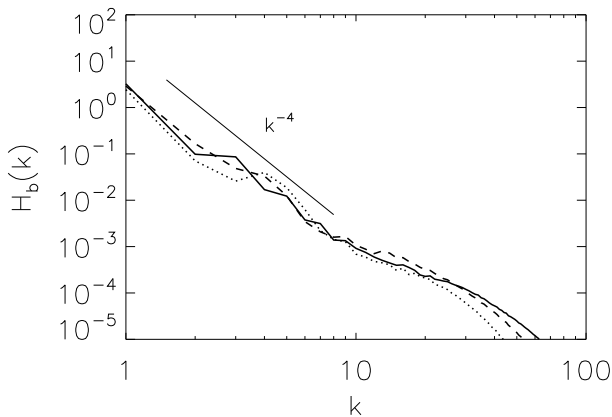}
\caption{\label{fig14} Magnetic helicity spectrum for three values of $R_m$:
          $R_m=2000$ (dotted), $R_m=5000$ (dashed), $R_m=10000$ (solid)
          for the ABC, $k_f=2$ flow. }
\end{figure}

The magnetic and kinetic helicity as a function of $R_m$ is shown in figure \ref{fig13}.
The last three points in this figure were obtained from the numerical simulations after roughly 
one diffusion time where the magnetic helicity was still slightly increasing with time.
To obtain a well converged averaged value many diffusion times
would be required that is not feasible with the present computational power.
Thus for these points the saturated value of the magnetic helicity is expected to be slightly higher than
what is shown in figure \ref{fig13}.
From the figure however one can see that the saturated value of the 
magnetic helicity  is slowly increasing with $R_m$ while hydrodynamic helicity
is approaching an asymptotic value.
As discussed in the introduction this positive value of magnetic helicity
implies that negative helicity has cascaded to the small scales where it 
was dissipated while positive helicity was concentrated at the large scales.
At the non-linear stage the flow is altered so that although
still helical (see fig. \ref{fig13}) it is less 
efficient at generating large scale helicity. 
The spectrum of the magnetic helicity is shown in \ref{fig14}. 
Unlike the kinematic stage in which negative helicity is concentrated in the small scales
and positive in the large, at saturation helicity
has the same sign
in both large and small scales. This implies that at the nonlinear stage some of the 
large scale helicity has been cascaded to the small scales. The transfer of
the large scale helicity to the small scales has been observed in MHD simulations
where a scale to scale transfer of the magnetic helicity was investigated \cite{hely_6}.
It is also noted that the helicity spectrum is very steep.  
If the magnetic field was fully helical
at small scales then it would be expected that the helicity spectrum 
would scale like $H_m\sim E_b/k\sim k^{-2}$, but in fig. \ref{fig14} the slope is much steeper
($H_m \sim k^{-4}$) implying that both signs of helicity are present in the small 
scales with the positive sign only slightly dominating.  

\begin{figure}
\includegraphics[width=6cm]{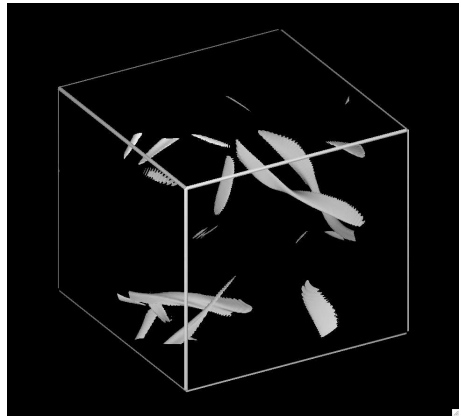}
\caption{\label{fig15} Magnetic energy density isosurfaces, for the Archontis Flow, $R_m=10^4$.
The surfaces correspond to 50\% of the maximum magnetic energy density or 10 times the average
magnetic energy.}
\end{figure}

The magnetic structures that appear for this flow can be seen in 
figure \ref{fig15} where  magnetic energy density isosurfaces are shown.
Again here energy is concentrated in long and flat structures (ribbons).
Although small scale structures are also present here they appear to be more
``organized" to form large scales patterns. Note the twisting of the ``ribbons"
that indicates the presence of helicity.

\section{Scalings at Saturation}

\begin{figure}
\includegraphics[width=8cm]{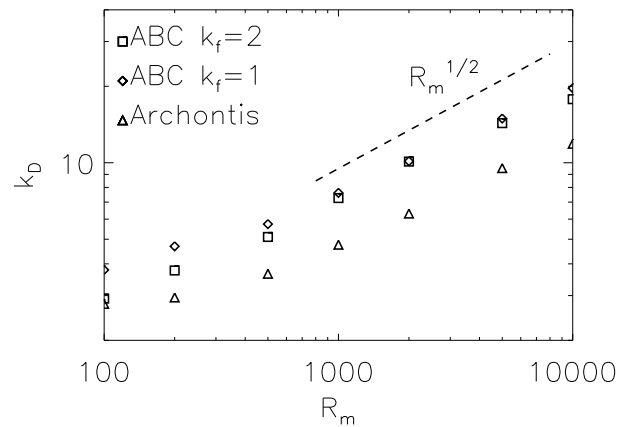}
\caption{\label{fig16} Magnetic energy dissipation wavenumber $k_{_D}$ as a 
function of $R_m$. }
\end{figure}

\begin{figure}
\includegraphics[width=8cm]{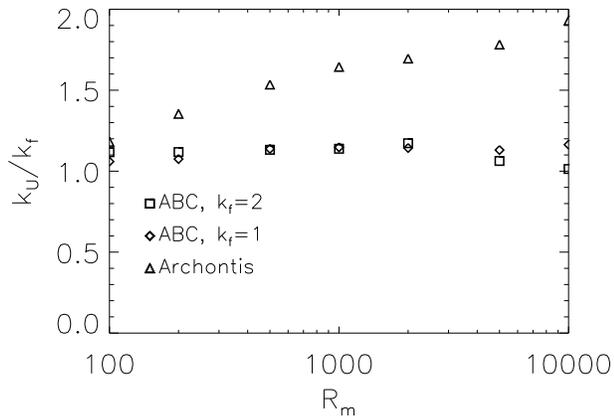}
\caption{\label{fig17} The kinetic energy dissipation wavenumber $k_{_U}$ as a function of $R_m$. }
\end{figure}

The three examined flows have some 
some common features that is worth further exploring. 
%
One of the fundamental relations from the kinematic theory that 
has been shown to hold for a variety of flows is that during the kinematic
stage the stretching term that scales like  $\sim UB/\ell_{_U}$
(where $U,B,\ell_{_U}$ are typical velocity, magnetic field, and velocity length scale respectably) 
is balanced by the magnetic diffusion term that scales like $\sim \eta B/\ell_{_D}^2$
(where $\ell_{_D}$ is the diffusion length scale).
This relation leads to the prediction that $\ell_{_D} \sim \ell_{_U}/\sqrt{R_m}$.
This relation continues to hold at the saturated stage. Fig \ref{fig16} shows 
the dissipation wave number $k_{_D}\sim 1/\ell_{_D}$ 
(defined as $k_{_D}^2=\langle (\nabla {\bf b})^2 \rangle /\langle {\bf b}^2 \rangle)$ 
as a function of $R_m$. For sufficiently large $R_m$ for all flows the 
dissipation wave number follows a scaling close to $k_{_D}\sim R_m^{1/2}$.
(A best fit gives a value of the exponent closer to $k_{_D}\sim R_m^{0.4}$.)
The velocity length scale on the other hand appears to depend weakly on 
the magnetic Reynolds number. In figure \ref{fig17} we plot the ratio of the velocity 
wavenumber  $k_{_U}^2=\langle (\nabla {\bf u})^2 \rangle /\langle {\bf u}^2 \rangle) \sim \ell_{_U}^{-2}$
to the forcing wavenumber $k_f$. $k_{_U}$ for the two ABC flows appears to be almost independent
of $R_m$ and close to $k_f$. For the Archontis flow the velocity wavenumber is weakly
increasing with $R_m$.  
Thus at the nonlinear stage both lengthscales $\ell_{_D}$ and $\ell_{_U}$ follow a similar
scaling with $R_m$ as they do in the the linear stage. Furthermore this scaling also explains why the Ohmic
dissipation that scales like $\epsilon_b \sim  B^2/R_m \ell_{_D}^2$ is of the same order with the viscous dissipation
that scales like $\epsilon_u \sim  U^2/\ell_{_U}^2$.

Besides the two length scales it is also interesting to investigate the effect of saturation
on the magnetic Reynolds number. An effective magnetic Reynolds number at saturation can be defined 
as $R_{m\, eff}=U_{nl}/k_{_U} \eta$, where
$U_{nl}=(2E_u)^{1/2}$ is the velocity amplitude at the nonlinear stage
and $k_{_U}$ the velocity wave number as defined in the previous paragraph.
If the dynamo saturates by decreasing the effective Reynolds number 
at the nonlinear stage then it would be expected that as $R_m$ is increased,
$R_{m\, eff}$ would approach the onset value $R_{mc}$.
\begin{figure}
\includegraphics[width=8cm]{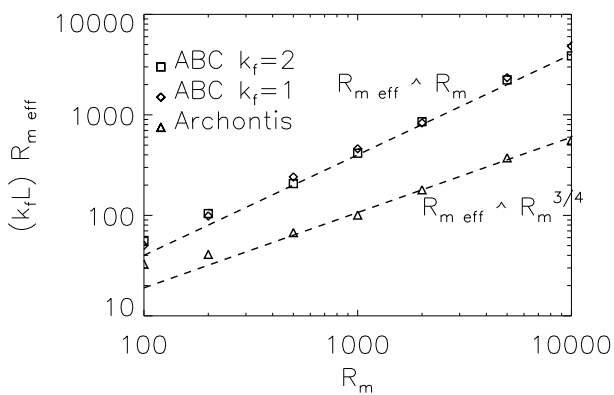}
\caption{\label{fig18} The effective magnetic Reynolds number $R_{m\,eff}$ (based on
the velocity amplitude and lengthscale at the nonlinear stage) as a function
of $R_m$ for the three different flows.}
\end{figure}
In figure \ref{fig18} we plot the effective Reynolds number as a function of 
$R_m$. For the ABC flows the relation between the two Reynolds numbers is very close to linear.
For the Archontis flow the $R_{m\,eff}$ is increasing with $R_m$ but weaker than linear
and closer to the scaling $R_{m\,eff}\sim R_m^{3/4}$.
Figure \ref{fig18} then implies that all relations described in this work for $R_m$ 
hold also for $R_{m\,eff}$ for the ABC flows, 
and for the Archontis flow provided the rescaling $R_{m\,eff}\sim R_m^{3/4}$ is taken into account.
Finally since $R_{m\,eff}$ is increasing with $R_m$ saturation 
is not due to a decrease of the amplitude of the velocity.  

\begin{figure}
\includegraphics[width=8cm]{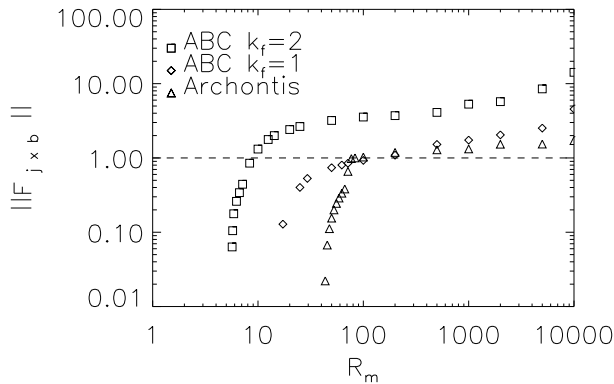}
\caption{\label{fig19} The amplitude of the Lorentz force for the three different flows as a function of $R_m$.}
\end{figure}

Although the magnetic field does not modify significantly the amplitude of the flow 
to saturate the dynamo it does modify its structure.
A measure of the strength of this modification is given by the amplitude of
the incompressible projection of the Lorentz force $F_{j\times b}={\bf b\cdot \nabla b} -\nabla P $.
Figure \ref{fig19} shows the amplitude $\|F_{j\times b}\|=\langle F_{j\times b}^2 \rangle^{1/2}$
as a function of $Rm$. What is observed is that for moderate values of $R_m$ ( $ 50< R_m <1000$ )
the amplitude of the Lorentz force is close to the forcing amplitude ($\|{\bf F}\|=1$).
As $R_m$ is further increased the amplitude of $F_{j\times b}$ seems to increase to larger values for the two ABC flows, 
while a weaker increase is observed for the Archontis flow. This suggests that at least for moderate $R_m$, 
$F_{j\times b}$ is in balance the body force $F$. 
At larger $R_m$ the effect of the Lorentz force is probably less organized due to the presence
of fluctuations and larger amplitude of $\|F_{j \times b}\|$ are needed to achieve a balance with $F$.

\begin{figure}
\includegraphics[width=8cm]{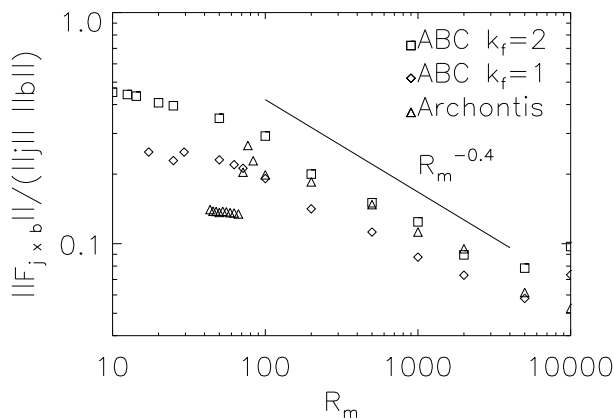}
\caption{\label{fig20} 
The amplitude of the Lorentz force normalized by $\|j\|\, \|b\|$ for the three different flows as a function of $R_m$. }
\end{figure}

One then needs to explain how a balance between a force generated by a small scale magnetic field ($\ell_D$)    
and a large scale force ($\ell_U$) is achieved.
In figure \ref{fig20} we show the relative amplitude of  $\|F_{j\times b}\|$
normalized by the amplitude of the current density $\|j\|=\langle {\nabla \times b}^2 \rangle^{1/2}$ and 
magnetic field amplitude $\|b\|=\langle {b}^2 \rangle^{1/2} $ as a function of the magnetic Reynolds number.
If the two fields ${\bf b}$ and ${\bf j=\nabla \times b}$ were randomly organized, the quantity 
$\|F_{j\times b}\| / (\|j\|\,\|b\|) $ should be of order one.
If however the two fields were aligned or were organized so that 
their cross product is a potential field the relative amplitude of
$\|F_{j\times b}\|$ can be a lot smaller.
In figure \ref{fig20} the relative amplitude of the Lorentz force 
appears to decrease as  a power-law for all flows with exponent close to
$\|F_{j\times b}\| \sim R_m^{-1/2} \|j\|\,\|b\|$.
(A best fit gives a value of the exponent closer to -0.4).
This can be understood in the following way. Saturation is expected to be achieved
when the Lorentz force is of the same order with the external forcing $F \sim F_{j\times b} = {\mathcal{O}}(1)$.
The amplitude of Lorentz force is also expected to scale as 
$F_{j\times b} \sim f \|j\| \cdot \|b\|$ where $f$ is a prefactor 
that depends on the alignment and structure of the two fields. Since the current 
density amplitude is controlled by the balance of stretching and diffusion
as seen in figure \ref{fig16}. We expect the scaling $\|j\| \sim \|b\| R_m^{1/2}/L$.
Equating the two relations we obtain that for an order one magnetic field
the prefactor $f$ must scale like $R_m^{-1/2}$.
This implies that the cross product of the two fields ${\bf j}$ and ${\bf b}$ must come 
closer and closer to a potential flow 
as $R_m$ is increased. This is achieved by forming the flat structures observed in figures
\ref{fig4},\ref{fig9} and \ref{fig15} where both the direction and the gradient of ${\bf j\times b}$
is in the direction perpendicular to their surface.
$F_{j\times b}$ then varies a much longer length scale than that of the magnetic field that is 
proportional to the curvature of these structures and is of the order of the forcing scale.
 
The observed deviations from this balance as $R_m$ is increased only reflect that $F_{j \times b}\sim F$
is not satisfied due to the presence of fluctuations that require $\|F_{j \times b}\| > \|F\|$
and thus stronger magnetic field to achieve saturation.
The behavior of these fluctuations due to magnetic instabilities thus 
seem to play an important role for the saturation at high $R_m$ and
need to be studied further.

\section{Summary and Discussion}

The nonlinear behavior of the dynamo instability was investigated in the infinite Prandtl number limit for three
different stationary flows.
In this limit large values of $R_m$ can be examined at moderate resolutions and this allowed us to investigate 
how the non-linear behavior of the dynamo scales with $R_m$.
What was shown for all flows is that the magnetic energy is increasing with $R_m$ at least logarithmically.
This increase was most pronounced for the ABC flow forced at $k_f=2$ for which an inverse cascade of magnetic
helicity was present.
The Ohmic and viscous dissipation rates on the other hand varied in behavior.
For the two $ABC$ flows they were approaching an asymptotic value as $R_m$ was
increased while for the Archontis flow were slowly decreasing $R_m$.
For all flows however the Ohmic and viscous dissipation rates although operating
at different scales were of the same order as a result of the balance between stretching and diffusion.

The saturation of the dynamo comes from a balance between the Lorentz force and the external body force
as their similar amplitude suggests. At saturation the magnetic field forms elongated flat structures
whose thickness scales like $R_m^{-1/2}$. The typical lengthscale along the other directions
was of the order of the forcing lengthscale. These structures were almost force-free 
with the solenoidal projection of the Lorentz force scaling like $F_{j\times b} \sim \|j\|\,\|b\|/\sqrt{R_m}$.
This decrease of $F_{j\times b}$ is the result of the organization of the magnetic field in these flat structures.
The cross product of the magnetic field with the current in these flat structures 
results in a field that is both pointing and varying in the direction perpendicular to the surface,
and thus is close to a potential field. In that way the magnetic field is successful in balancing 
the external force that differs both in amplitude (compared to $B^2/l_{_D}$) and length-scale. 
 
Another common feature that all flows had was that the generated structures were unstable to small
scale fluctuations that resulted in a chaotic behavior. Note that these fluctuations are due to 
magnetic instabilities and their origin is probably related to reconnection events.   
However, these chaotic fluctuations have not resulted in universal  spectra.
The observed slopes of the spectra varied from positive
($ABC k_f=1$ flow) to almost flat spectrum (Archontis flow) to negative slope ($ABC k_f=2$ flow). 
Thus if these instabilities lead to universal behavior it must happen for even larger magnetic 
Reynolds numbers.

Finally, a comment needs to be made for the applicability of these results to realistic flows.
It is the author's belief that although the dynamo has been investigated in a somehow idealistic limit some insight can be gained.
In a realistic flow there exist both large sale structured flows forced by a thermal or other instability
and turbulent small scale fluctuations as a result of a turbulent cascade. Although the turbulent 
fluctuations result in the fastest growing modes in a large Prandtl number dynamo flow
at saturation the magnetic field configuration must be such that the Lorentz
force prevents further magnetic field amplification due to both the turbulent fluctuations, and the large forcing scales.
The results in this work help in understanding the latter behavior.
In particular this work gives a clear example of how small scale fields 
can organize themselves to come in balance  with a large scale forcing.


\end{document}